\documentclass[aps,prd,showkeys,showpacs,amssymb,cite,
amsfonts,epsf,preprintnumbers,nofootinbib,superscriptaddress]{revtex4}

\usepackage[dvips]{graphicx}
\usepackage{bm,latexsym,amsmath,amssymb,amsfonts}
\usepackage[usenames,dvipsnames]{color}
\usepackage[colorlinks=true,linkcolor=blue]{hyperref}
\usepackage{color}
\usepackage{soul}
\usepackage{epsfig}
 
\definecolor{mypink1}{rgb}{0.858, 0.188, 0.478}
\definecolor{mypink2}{RGB}{219, 48, 122}
\definecolor{mypink3}{cmyk}{0, 0.7808, 0.4429, 0.1412}
\definecolor{mygray}{gray}{0.6}
\definecolor{pptbg}{rgb}{0.961,0.945,0.863}

\newcommand{\be}[1]{\begin{equation} \label{#1}}
\newcommand{\ee}{\end{equation}}
\newcommand{\bea}{\begin{eqnarray}}
\newcommand{\eea}{\end{eqnarray}}
\newcommand{\ba}{\begin{array}}
\newcommand{\ea}{\end{array}}
\newcommand{\bel}{\begin{align}}
\newcommand{\eel}{\end{align}}
\newcommand{\nn}{\nonumber}
\newcommand{\nnl}{\nonumber \\}

\begin{document}

\title{Entropy of a Wormhole from the Constituent}

\author{Hyeong-Chan Kim}
\email{hckim@ut.ac.kr}
\affiliation{School of Liberal Arts and Sciences, Korea National University of Transportation, Chungju 27469, Korea}
%
\begin{abstract}

We obtain the entropy of a static spherically symmetric wormhole from the entropy of matter consisting it.
Some exact wormhole solutions are analyzed in detail and their entropies are obtained. 
We display the asymptotic forms of the entropy for solutions having regular asymptotic regions in which matter spreads to infinity.
In the analysis, we find that the combination of ADM masses $\mathcal{M}\equiv \alpha_+ M_+ + \alpha_- M_-$ plays an important role in thermodynamics of a wormhole,
 where $\displaystyle \alpha_\pm \equiv \lim_{r\to \infty} \sqrt{-g_{tt}}$ in the upper/lower half, respectively.
\end{abstract}
\maketitle
\section{Introduction}
A wormhole is an interesting object which connects distant asymptotic regions through a handle, of which length could be short enough irrespective of the distance outside.
It appears as a solution of Einstein equation, however, 
its unique geometry obscures various previously accepted concepts in general relativity presenting an opportunity to reconsider them.  
To form a wormhole throat in general relativity, `exotic' or `phantom' matter is necessary~\cite{Morris:1988cz,Raychaudhuri:1953yv}.
One may construct a time machine~\cite{Morris:1988tu,Frolov:1990si} and
may have `charge without charge'~\cite{Misner:1957mt}.
The ADM mass measured in one side of the wormhole
may not be identical to that in the other side~\cite{Frolov:1990si,Kim:2019elt}.
For a comprehensive review of the topic, please see the book~\cite{Visser:1995cc}.
Variations of the wormhole models were also developed. 
A phantom free models~\cite{Bronnikov:2018uje}, a charged wormhole~\cite{Kim:2001ri},
and wormholes in modified theories of gravity
~\cite{Mehdizadeh:2015jra,Zangeneh:2015jda,Mazharimousavi:2016npo,Nandi:1997mx,Eiroa:2008hv} were studied.
Stability of a wormhole were also studied~\cite{Bronnikov:2013coa,Novikov:2012uj}.
The possibility of observing a wormhole from the movement of stars has been discussed recently~\cite{Dai:2019mse,Krasnikov:2019}.
Wormholes have also drawn interests
as a test-bed to understand quantum entanglement~\cite{Maldacena:2013xja,Maldacena:2018gjk}.
The study of wormholes as physical entities in understanding the quantum-gravity era
is getting more important than ever at this juncture.

 Thermodynamical laws play crucial roles in describing a macroscopic system. 
Thermodynamic properties provide a star with matter equation of state composing it. 
As for a black hole, its main characteristics are the same as those of  thermodynamics~\cite{BCH}. 
Even in the presence of gravity, the ordinary thermodynamical laws work well in a simply connected spacetime having one asymptotic region.
 Asymptotic temperature here serves as the key parameter describing the thermal equilibrium~\cite{Tolman1930}. 
However, when a spacetime has various asymptotic regions or is connected multiply, the ordinary thermodynamic laws do not apply as shown in Ref.~\cite{Kim:2019elt}.  
Typical examples having the properties are the wormhole spacetimes, such as a wormhole between universes and  a wormhole within a universe.
There are a few reasons.
i) The total energy of a wormhole is difficult to identify. 
ii) The entropy of a wormhole is not yet known. 
iii) The asymptotic temperature of a wormhole fails to characterize its thermal equilibrium. 
For an inter-universe wormhole, these difficulties originate from the fact that there are many asymptotic regions.
Because of that, the ADM mass of a wormhole measured in one asymptotic region may not be identical to that of the other. 
For an intra-universe wormhole, 
the ADM mass is ill-defined even though there is one asymptotic region. 
The time-like Killing vector field may not exist even though the spacetime is locally static everywhere.
In this work, we are mainly interested in inter-universe wormholes. 
For an intra-universe one, we discuss briefly in the last section.

\vspace{.1cm}
Let us begin with the metric of an inter-universe wormhole spacetime.
Assuming spherical symmetry, the geometry of the upper/lower half with respect to the wormhole throat can be described by the line element,
\be{metric}
ds_\pm^2 = - f_\pm(r) dt^2 + \frac{1}{1-2m_\pm(r)/r} dr^2 + r^2 d\theta^2 + r^2 \sin^2\theta d\phi^2, \qquad r\geq b,
\ee
where the $\pm$ represents the upper/lower half and $b$ represents the size of wormhole throat. 
The upper and the lower geometries should be adjoined at $r=b$ smoothly. 
Therefore, there are two spheres with the same radius $r (> b)$ each in the upper and the lower halves.
The mass function $m_\pm(r)$ is
\be{gr}
 m_\pm(r) = \frac{b}{2} + 4\pi \int_b^r r'^2 \rho_\pm(r') dr',
\ee
where $\rho_\pm(r)$ represents the matter density composing the wormhole.
Here, we have imposed the mass function to have the value $b/2$ at the wormhole throat.
The metrical function $f_\pm(r)$ and $m_\pm(r)$ will be determined by the Einstein equation.

Let the mouths of the wormhole be located at $r_\pm$ for the upper/lower half so that the matter be confined in $b \leq r \leq r_\pm$. 
The geometry outside of the mouths is described by the Schwarzschild metric, 
\be{metric:out}
ds_\pm^2 = - \alpha_\pm^2 \Big(1-\frac{2M_\pm}{r}\Big) dt^2 + \frac{1}{1-2M_\pm/r} dr^2 + r^2 d\Omega^2, \quad r\geq r_\pm,
\ee
where $M_\pm$ represents the ADM mass of the wormhole measured in the upper/lower half and $\displaystyle \alpha_\pm^2 \equiv \lim _{r\to\infty} f_\pm(r)$.

The thermodynamic equilibrium condition in the presence of a wormhole was studied in Ref.~\cite{Kim:2019elt}.
There, a generalized temperature was proposed which identifies the thermal equilibrium between systems.
In the work, the authors have assumed that the entropy of a wormhole comes from the matter consisting it. 
%
Given the asymptotic temperature $T_{\rm asym}^{\pm}$ for the wormhole, the generalized temperature becomes
\be{Tgen1} 
\mathfrak{T} \equiv \alpha_+ T_{\rm asym}^+ = \alpha_- T_{\rm asym}^- .
\ee
There are a few merits in the use of the generalized temperature.
i) The generalized temperature is unique everywhere even if the space has many asymptotic regions.  
ii) The generalized temperature is well-defined even for an intra-universe wormhole. 
iii) As shown in the work, the generalized temperature can be obtained from the physical data localized at a point.
This is in contrast to asymptotic temperature, which requires data at infinity also. 

The matter consisting the wormhole is anisotropic in general. 
The stress tensor for an anisotropic fluid compatible with the spherical symmetry must be diagonal and the pressure must be isotropic in its angular directions.
In the present work, we consider a linear equation of state that the radial and the angular pressures to be proportional to the density:
\be{eos}
p_1 = w_1 \rho, \qquad p_2 = w_2 \rho.
\ee
The entropy, then, can be obtained by integrating the entropy density $s$ over spatial volume,
\be{S1}
S_{\rm WH} = S^+(r_+) + S^-(r_-); \qquad
S^\pm(r) \equiv \int_{b}^{r} d\Sigma_\pm s_\pm^a n_a 
	 = \int_{b}^{r} L_\pm(r') dr'  ,
\ee
where  $S^\pm(r)$ represents the total entropy of matter in $r$ in the upper/lower half.
 The Lagrangian-like term is
\be{L1}
L_\pm(r) \equiv
\frac{ 4\pi r^2 s_\pm(r)}{\sqrt{1- 2m_\pm(r)/r}}  ,
\ee
where $s_\pm(r) $ denotes the entropy density of matter at $r$ in the upper/lower half.
The entropy density for an anisotropic matter was given in Ref.~\cite{Kim:2019ygw}.

In Sec.~\ref{sec:II}, we reduce the Einstein equation for a wormhole spacetime to a first order autonomous equation in a two-dimensional plane.   
We also show a few important properties of the solution curve on the plane.
In Sec.~\ref{sec:III}, we obtain the general entropy formula for a wormhole by integrating the entropy density of matter over volume. 
In Sec.~\ref{sec:IV}, we calculate entropies of a few exact wormhole solutions.  We also study some properties of the exact solutions which were not mentioned in the previous literatures. 
In Sec.~\ref{sec:V}, we summarize the results.

\section{Einstein equation and Wormhole solutions} \label{sec:II}

Most of the results in this section are brief summaries of Refs.~\cite{Kim:2019ojs,Kim:2017hem}, up to some tweaks to the notation to simplify writing and discussion.

The outside geometry for $r\geq r_\pm$ is nothing but the Schwarzschild solution given in Eq.~\eqref{metric:out} with ADM mass $M_\pm$. 
The Einstein equation describing the handle part of a spherically symmetric wormhole is based on the metric~\eqref{metric}.
It gives, in addition to Eq.~\eqref{gr}, 
\bea
\label{Eineq}
\frac{f_\pm'}{f_\pm}&=&\frac{2(m_\pm+4\pi r^3p_{\pm 1})}{r^2\left(1-2m_\pm/r\right)}.
\eea
This equations and $G^\theta_{~\theta}=8\pi T^\theta_{~\theta}$
give the modified Tolman-Oppenheimer-Volkhoff (TOV) equation:
\bea
\label{TOVA}
\frac{dp_{\pm1} }{dr}=-\frac{\rho_\pm+p_{\pm 1} }{r(r-2m_\pm )}\left(m_\pm +4\pi r^3p_{\pm 1}\right)+\frac{2}{r}(p_{\pm 2}-p_{\pm 1}).
\eea
To simplify the differential equation, we introduce two scale invariant variables,
\be{uv:def}
u = \frac{2m_\pm(r)}{r}, \qquad x = v_b^{-1}\frac{dm_\pm(r)}{dr} = \frac{4\pi  r^2 \rho_\pm(r)}{v_b} ; \qquad v_b = -\frac{1}{2w_1}.
\ee
An interesting finding on this scale invariant variables is that the throat values are determined to be
$$
(u_{\rm throat}, x_{\rm throat}) = (1, 1).
$$

Using the variables, the TOV equation~\eqref{TOVA} can be reduced to a first order autonomous equation,
\be{TOV}
\frac{du}{dx} = \frac1{F(u,x)} \equiv 2q \frac{(1-u)[ (1-q^{-1}) x-u]}{x[x-1 + s(1-u)]}, 
\qquad x \equiv \frac{v}{v_b},
\ee
and a radial equation  
\be{xi:r}
\frac{dy}{y} = \frac{dr}{r} = \frac{du}{(1-q^{-1}) x-u}= 2q\frac{1-u }{x-1 + s(1-u)} \frac{dx}{x} , 
\qquad y \equiv \frac{r}{b}.
\ee
Here $q$ and $s$ are constants introduced for notational convenience and are determined by the equation of states as follows:
\be{vb}
q = \frac{w_1}{1+w_1}, \qquad
s = 1+\frac{4w_2}{1+w_1} .
\ee
For a wormhole throat to exist as shown in Ref.~\cite{Kim:2019ojs}, $q$ must be positive, not 1, that is, $w_1< -1$ or $w_1 > 0$.
The signature of the density is always the same as $v_b$, so $x$ must be a non-negative number. 
In this work, we do not put $\pm$ indices in $u$ and $x$ variables because it is self-explanatory which half (upper or lower) the given point $(u,x)$ belongs to, as will be shown soon.  
Rather, we use $(u_\pm,x_\pm)$ to denote boundary values at the mouths of the wormhole. 
Given a set of boundary data $(u_+,x_+)$, the autonomous equation~\eqref{TOV} presents a unique solution curves $C$ on the $(u,x)$ plane. 
Let us summarize a few important properties of solution curves from the wormhole point of view.
\begin{enumerate}
\item Any solution curve having a wormhole throat fails to pass the two lines $u=1$ and $v=0$. 
\item A wormhole throat forms at $(u,x)=(1, 1)$. 
Around the throat, the solution curve $C_\kappa$ and the corresponding radial solution to zeroth order is 
\be{0th}
C_\kappa:
\kappa(1-u) = \big(x-u\big)^2, \qquad y = 1+q (1-u).
\ee 
Here, $\kappa$ denotes the amount of asymmetry\footnote{The variable $\kappa$ denotes $(\kappa v_b^2)^{-1}$ in the  paper~\cite{Kim:2019ojs}. This notational change make it easier to represent the symmetric limit.} between the upper and the lower halves. 
We divide $C_\kappa$ into two pieces $+~ (x>u)$ and $- ~(x< u)$,  according to the signature of $x$ at the just outside of the throat.
We call them the upper and the lower halves of the curve, respectively. 
When $\kappa = 0$, the symmetric solution $u=x$ appears. 
For a given equation of state, solution curves are uniquely characterized by $\kappa$.   
\item The radius increases as the point on the curve moves away from the throat value.
\item Exact solution curves were found when $s=0$ and $s=-1$. 
In other cases, numerical solution curves were described.
\end{enumerate}
Given a point $(u,x)$, one can clearly determine which piece of $C_\kappa$ the given point belongs to.
So one does not need ``$\pm$'' in $(u,x)$ because it is self-evident.

The undetermined metrical function in Eq.~\eqref{metric} becomes
\bea
f_\pm (r) = f_b x^{-2q }  y^{s-1} , \label{f pm}
\eea
where $f_b$ is the value of the metric function at the throat.

\section{Entropy of matter in a wormhole}\label{sec:III}

The relation between the entropy density and the energy density must be dependent only on the physical properties of matter itself. 
The entropy density of an anisotropic matter is, in Ref.~\cite{Kim:2019ygw},  
\be{sA}
s_\pm (v, r) = 
 (1+w_1) \sigma\left|\frac{\rho_{b'}}{\sigma} \right|^{\frac{1}{1+w_1}}  \left(\frac{\rho_\pm}{\rho_{b'}} \right)^{\frac{1}{1+w_1}} \left(\frac{r}{b'} \right)^{\frac{2(w_2-w_1)}{1+w_1}}
	= (1+w_1) \sigma
		\left|\frac{\rho_{b'}}{\sigma} \right|^{1-q}  
		\left(\frac{x}{x_ {b'}}\right)^{1-q} \left(\frac{r}{b'} \right)^{\frac{s-5}{2}}  ,
\ee
where $\sigma$ is a constant which depends on the species of the matter. 
Here, we introduce an arbitrary radial scale $b'$ and density scale $\rho_{b'}$ to make the dimension of $\sigma$ to be the same for all $w_i$ and to remove the ambiguity on the signatures in $\rho_\pm$. 
From now on in this work, we choose the scales to be those at the wormhole throat: $x_{b'} = x_b=1$ and $b' = b$.

For notational simplicity, we introduce $\sigma' \equiv \sigma \left|\rho_{b}/\sigma \right|^{1/(1+w_1)}
= \sigma \left| v_{b}/4\pi b^2 \sigma \right|^{1-q} $.
Using the entropy density,
the Lagrangian-like term in Eq.~\eqref{L1} becomes 
\be{L}
L_\pm(r) \equiv
\frac{ 4\pi (1+w_1) \sigma'  r^2}{\sqrt{1- 2m_\pm(r)/r}} 
  	\left( \frac{\rho_\pm(r)}{\rho_{b} }\right)^{1-q}
	\left(\frac{ r}{b}\right)^{\frac{2(w_2-w_1)}{1+w_1}} 
=\frac{ 2\pi  \sigma' b^2}{-qv_b} 
  	\frac{x^{1-q} y^{\frac{s-1}{2}} }{\sqrt{1-u}},
\ee
where we use $x$ and $y$ in Eqs.~\eqref{TOV} and \eqref{xi:r}.
For $L_\pm$  to be positive, $- \sigma'/v_b= 2w_1\sigma' > 0$.
Let the values of $(u,x)$ at $r= r_\pm$ to be $(u_\pm, x_\pm)$.
Note that the variational equation of $L_\pm$ with respect to $m(r)$ presents the TOV equation~\eqref{TOV}, which respects the so-called maximum entropy principle~\cite{Sorkin:1981wd}.
The total entropy can be obtained by integrating $L_\pm$ as in Ref.~\cite{Kim:2019ygw},
\be{Sa}
S^\pm \equiv \int_b^{r_\pm} L_\pm(r)dr =  S(u_\pm,x_\pm,y_\pm) - S_\pm(b) ,  
\ee
where $S(u,x,y)$ is an entropy function given by 
\bea \label{Sr-2} 
S(u,x,y) &=& \frac{r}2\frac{2m/r+ 8\pi r^2 p_{1}}
	{\rho+ p_{1} + 2p_{2} } 
		\frac{s (r)}{(1- 2m/r)^{1/2}} 
= \frac{4\pi b^3\sigma'}{v_{b}(s+1)} 
	\frac{(u-x) x^{-q}}{\sqrt{1-u}} 
	y^{\frac{s+1}{2} } .
\eea
Note that $S(u,x,y)$ changes its signature at the line $u= x $. 
This signature change does not imply that a physical entropy can take a negative value because it is defined by the difference in Eq.~\eqref{Sa}. 

Let us consider the $s \neq -1$ case. 
We deal the $s=-1$ case in a separate Sec.~\ref{sec:IVB}.  
Note that at the wormhole throat, by using Eq.~\eqref{0th},
\bea
S_\pm(b) = \lim_{r\to b} S(u,x,r) 
&=& \frac{{ 4\pi b^3\sigma' }  }{ v_{b} (s+1)} 
	\lim_{r\to b} \frac{-(1- u)+1 -x}{(1-u)^{1/2}}
	x^{-q} y^{\frac{s+1}{2}}   \nn \\
 &=&  \frac{ 4\pi b^3 \sigma'  \sqrt{\kappa} }{ v_{b} (s+1)} 
	  \lim_{x\to 1_{\pm}} \frac{1-x}{|1-x|}
		 x^{-q} y^{\frac{s+1}{2}} \nnl
 &=&   \mp \frac{4\pi b'^3\sigma' \sqrt{\kappa}   }
 	{v_b (s+1) }  
	.
	\label{S:throat}
\eea
Here $\pm=\mbox{sign}(x -1)$ represents the upper/lower half.
Putting the result into Eq.~\eqref{Sa}, the total entropy of the throat including both sides becomes
\bea \label{S WH}
S_{\rm WH} &=& S^+(u_+,x_+,y_+) + S^-(u_-,x_-,y_-); \nnl
S^\pm(u,x,y) &=& \frac{4\pi b^3 \sigma'}{ v_b (s+1)}  
	\left[ 
	\frac{u -x }{\sqrt{1-u}} x^{-q } 
		y^{\frac{s+1}{2} }  \mp \sqrt{\kappa} \right] , \quad s\neq -1.  
\eea
Here $u_\pm$, $x_\pm$, and $y_\pm$ represent the corresponding values at the mouths of the upper/lower half. 
Note that the $s\to -1$ limit is well-defined when $\frac{u -x  }{\sqrt{1-u}} x^{-q }=\sqrt{\kappa} $. 
In fact, this relation presents an exact solution for the case.
 To get the entropy of the system by using the limit, one need to know solutions of $(u,x,y)$ to the first order of $s+1$. 
Taking the limit $s\to -1$ after putting the first order solution to Eq.~\eqref{S WH}, one may obtain the entropy with $s=-1$. 
In this work, we rather take a simpler way than the limiting process.
We integrate of the entropy density directly,  
which calculation is shown in Sec.~\ref{sec:IVB}.

\section{Entropy of some exact wormhole solutions} \label{sec:IV}
In this section, we calculate the entropies of exact wormhole solutions and near throat solutions.
We also calculate entropies of asymptotically flat wormholes in which matter spreads throughout the universe.
The solutions we deal in this section hold only in the matter-filled region.
The region outside will be described by the spherically symmetric vacuum Schwarzschild solution. 

\subsection{ The entropy of a wormhole with matter equation of state $s=0$}
\label{sec:IVA}

In this case, the equation of state satisfies $1+w_1+4w_2=0$.
We consider the geometry of a wormhole handle, filled with matter, between the two mouths.
The solution curve describing the handle located at $x_-\leq x \leq x_+$ is given by~\cite{Kim:2019ojs}
\be{sol1}
1-u = \frac{(1-x)^2}{1+ \frac{2q}{1-2q} x +\Big(\kappa -\frac{1}{1-2q}\Big) 
	x^{2q} }.
\ee
They are characterized by $\kappa$, the amount of asymmetry in Eq.~\eqref{0th}. 
A radial scale is given by the throat size $b$.
Now, once the position of mouths $x_\pm$ are given, every data of the wormhole solution are fixed. 
Summarizing, we have four independent data $(\kappa, b, x_+, x_-)$.
The radius satisfies 
\be{y:sol1}
y \equiv \frac{r}{b} =  {\kappa^{-1}} x^{-2q} 
	\left[ 1+ \frac{2q}{1-2q} x 
	+\Big(\kappa -\frac{1}{1-2q}\Big)  x^{2q}  \right].
\ee
For the lower half, $y$ varies from $1$ to infinity as $x$ decreases to zero. 
For the upper half, $y$ monotonically increases from $1$ to a finite value $ 1+[\kappa(2q-1)]^{-1}$ for $q> 1/2 $ with $x$.
For $0< q<1/2$, it goes to infinity as $x\to \infty$.
The mass function becomes
\be{m:sol1}
m(x) = \frac{b}{2\kappa}\left[ \frac{2(1-q)}{1-2q} x^{1-2q} - x^{2-2q} 
		+ \kappa + \frac{1}{2q-1}\right].
\ee
When $0< q< 1/2$, the mass approaches a finite value as $x\to 0$ ($r_- \to \infty$).
On the other hand, it goes to infinity as $x\to \infty$ ($r_+\to \infty$). 
When $q> 1$, the mass approaches a finite value as $x\to \infty$ but diverges at $x= 0$.  
%
The metric now can be written as, by using $x$ as a radial coordinate, 
$$
ds^2 = - \frac{f_b}{ x^{2q}y(x)} dt^2 
+\frac{4b^2q^2}{\kappa}\frac{ y(x) }{x^{2+2q}  } dx^2 + b^2 y(x)^2 d\Omega^2, \quad x_- \leq x \leq x_+.
$$

\begin{figure}[hbtp]
\begin{center}
\includegraphics[width=0.55\textwidth]{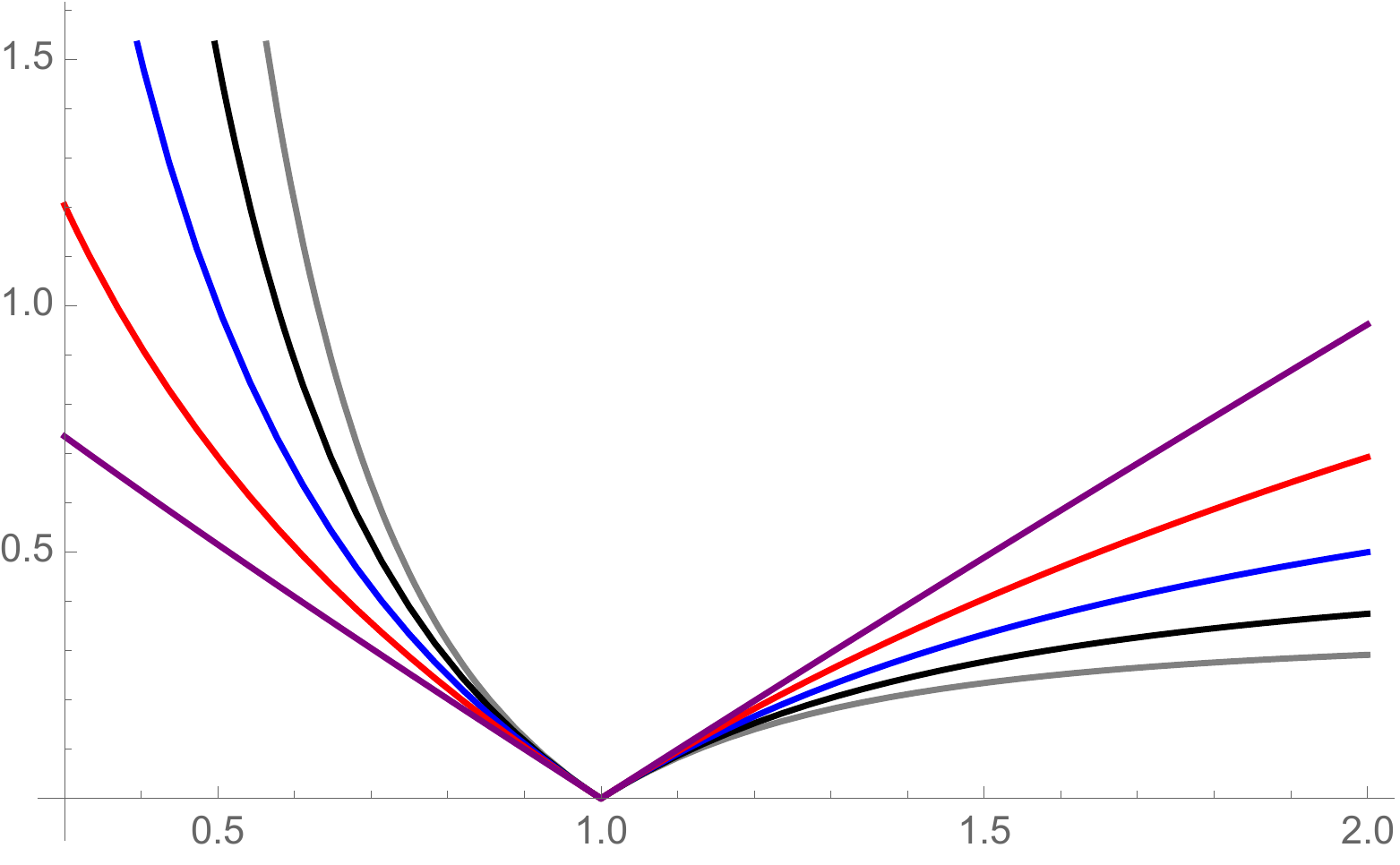}
\put( -1,10) {\large $x$}  \put(-280,170) { $S^\pm$}
\caption{The entropy inside each mouth $\frac{-v_b \sqrt{\kappa} }{4\pi b^3\sigma'}S^\pm$. 
Here, $\displaystyle q =2, 3/2, 1,0.50005, $ and $0.05$, respectively for gray, black, blue, red, and purple curves.
The throat is located at $x=1$. 
$S^\pm$ is plotted in the region $x \gtrless 1$, respectively. 
}
\label{fig:tm}
\end{center}
\end{figure} 
The entropy~\eqref{S WH} for the upper and the lower halves become
\bea \label{Sr-3} 
S^\pm &=& \frac{4\pi b^3\sigma' \sqrt{\kappa} }{v_b }  \left[\left(-(1-u_\pm)+1 -x_\pm\right)
	\frac{x_\pm^{-q} y_\pm }{\sqrt{\kappa(1-u_\pm)y_\pm}} 
		 \pm 1\right]\nnl
&=& \pm \frac{4\pi b^3\sigma' }{v_b\sqrt{\kappa} } \frac{1-
 	x_\pm ^{1-2q} 
 	 }{1-2q} \geq 0
 .
\eea
$S^\pm=0$ at the throat by definition. 
When  $q>1/2$ ($|w_1| > 1$), $S^-$ diverges at $x_-=0$.  
The value of $S^+$ remains finite for all $x_+ \geq 1$.
On the other hand when $0 < q < 1/2$ ($0<w_1<1$), $S^-$ remains finite for all $0\leq x_- \leq 1$ but $S^+$ diverges as $x_+\to \infty$.
This behavior is consistent with the mass form in Eq.~\eqref{m:sol1}.
The distribution of the entropy around the throat is closer to a symmetrical form with a smaller $q$.
On the other hand, as $w_1$ gets closer to $-1_{-}$ ($q\to \infty$), the distribution becomes more asymmetric.

Given the values $x_\pm$ at the mouths, the entropy of the wormhole now becomes
\be{SWH:t=0}
S_{\rm WH}  = S^+ + S^- = \frac{4\pi b^3\sigma' }{-v_b\sqrt{\kappa}}
	\times
	 \frac{x_+^{1-2q} -x_-^{1-2q}}{1-2q}.
\ee
The entropy for the case $q=1/2$ can be obtained from the $q\to 1/2$ limit of Eq.~\eqref{SWH:t=0}.

\subsection{ The entropy of a wormhole with matter $s=-1$}
\label{sec:IVB}

In this case, $1+w_1+2w_2=0$. 
As in Ref.~\cite{Kim:2019ojs}, a solution curve $C_\kappa$ for a general wormhole handle is given by 
\be{sol}
\kappa(1-u) = \left(x -u\right)^2 x^{-2q } . 
\ee
The symmetric solution $x=u$ appears when $\kappa=0$. 
Using $x -u = x-1 +(1-u),$ 
$u$ can be expressed as a function of $x$: 
\be{sol2} 
1-u = 1- x + \frac{\kappa}{2} x^{2q}\mp \sqrt{\kappa} x^q \sqrt{ \frac{\kappa}{4} x^{2q} 
		+ 1- x} 
= \left(\frac{\sqrt{\kappa} }{2} x^q \mp \sqrt{ \frac{\kappa}{4} x^{2q} + 1- x} \right)^2.
\ee
Here, the solution curve $C_{\kappa}$ is divided into two pieces, the Left (+) and the Right ($-$), because $u$ is double valued with respect to $x$. 
This division is different from the previous upper/lower half with respect to the throat.
The behaviors of the solution curves are divided into two classes depending on the existence of a real-valued root of the equation,
\be{xM}
 \frac{\kappa}{4} x_M^{2q} + 1- x_M =0.
\ee
When $\kappa > 2q^{-1}(1-1/2q)^{2q-1}$, there is no real root.
In this case, the Left and the Right curves do not meet each other and represent distinct solution curves.
Each curve presents a unique solution of Einstein equation.
The Right curve contains a wormhole throat at $x=1$. 
We are not interested in the Left curve because it does not contain a wormhole throat.
When $\kappa \leq 2q^{-1}(1-1/2q)^{2q-1}$, the Left and the Right curves are connected smoothly at $(u,x) =(2-x_M, x_M )$ to form one solution curve $C_\kappa$.
There, $x$ takes its maximum.
The Left curve satisfies $u \leq u_M<1$ and stays away from the throat. 
The Right curve contains a wormhole throat at $u=1$.

Let us first consider the case with $\kappa > 2q^{-1}(1-1/2q)^{2q-1}$ and $q> 1/2$.
The Right curve describes the wormhole throat, where $x$ varies from zero to infinity. 
Because $\kappa$ cannot vanish, there is no symmetric solution.
The radius of the asymmetric solution satisfies, from Eq.~\eqref{xi:r}, 
\be{r:sol1-1}
\frac{r}{b} = y = x^{-q} \times
 \exp \left|\mathfrak{I}_{q,\kappa} (x) \right| ,
 \ee
where 
\be{I}
\mathfrak{I}_{q,\kappa} (x) \equiv \frac{q \sqrt{\kappa}}{2} 
	\int_1^x \frac{ x'^{q-1}dx'}
	 { \sqrt{\frac{\kappa}{4}
			x'^{2q}+1-x' }} .
\ee
As $x\to 0$, it approaches a constant value forming an asymptotic region. 
For large $x$, $\exp\left|\mathfrak{I}_{q,\kappa} (x)\right|$ behaves as $e^c\times x^q $ with positive $c$.
Therefore, $r \to b e^c$ as $x \to \infty$. 
The mass takes the form,
\be{m:sol1-1}
m(x) = \frac{byu}{2} = \frac{b}{2}
  \left[x^{1-q}-\frac{\kappa x^q}{2} +\sqrt{\kappa}
 	\sqrt{\frac{\kappa}4 x^{2q}+ 1-x}
	 \, \right] \exp \left|\mathfrak{I}_{q,\kappa} (x) \right|  .
\ee
When $q> 1$, the mass function diverges in the both limits $x\to \infty$ and $x\to 0$. 
On the other hand, when $0<q<1$, it goes to infinity or approaches a finite value for $x\to \infty$ or $x\to 0$, respectively.
The metric function $f_\pm(r)$ becomes
\be{f:sol2-1}
f_\pm(r) = f_b 
  \exp \big(-2\left|\mathfrak{I}_{q,\kappa} (x)\right| \big).
\ee
Here, $\pm$ represents both ends of the solution curve.

To find the entropy, one may resort to the formula~\eqref{S WH}.
To get the $s\to -1$ limit, one need to find a first order solution of Eq.~\eqref{TOV} to the order $O(s+1)$.
Then, one can take the limit.  
However, the calculation is quite cumbersome. 
Here, to get the total entropy, $S_{\rm WH} = S^-+S^+$, we directly integrate $L^\pm$, 
\bea
L_\pm &=& \frac{ 2\pi \sigma' b^2}{-q v_b}  
  	\frac{x^{1-q}y^{-1}}{\sqrt{1-u}}   
= \frac{2\pi b^2 \sigma'\sqrt{\kappa}}{-qv_b} \frac{x}{|x-u| y }.
\eea
The integration becomes, by using Eqs.~\eqref{sol} and \eqref{sol2}, 
\be{S:sol2}
S^\pm=\int_b^{r_\pm} L_\pm dr = \frac{4\pi b^3 \sigma'}{-v_b} \left|K_{q,\kappa}(x_\pm)\right| 
\ee
where 
\be{Kfn1}
K_{q,\kappa} (x) \equiv \int_1^{x} \frac{x'^{-q}
	 }{ \sqrt{\frac{\kappa}{4}x'^{2q}+1-x' }} dx'.
\ee
\begin{figure}[hbtp]
\begin{center}
\begin{tabular}{cc}
\includegraphics[width=0.4\textwidth]{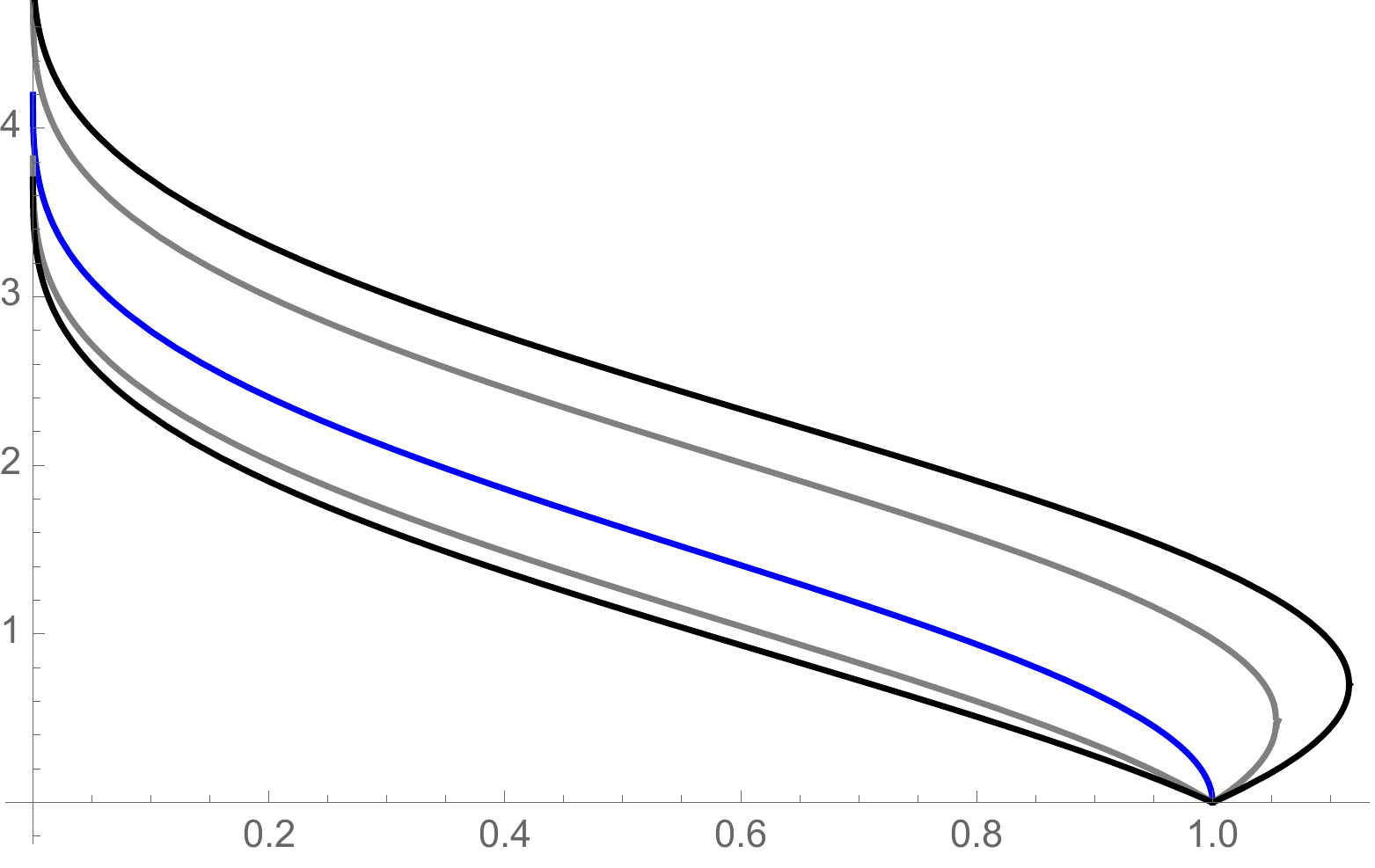}& 
\quad
\includegraphics[width=0.4\textwidth]{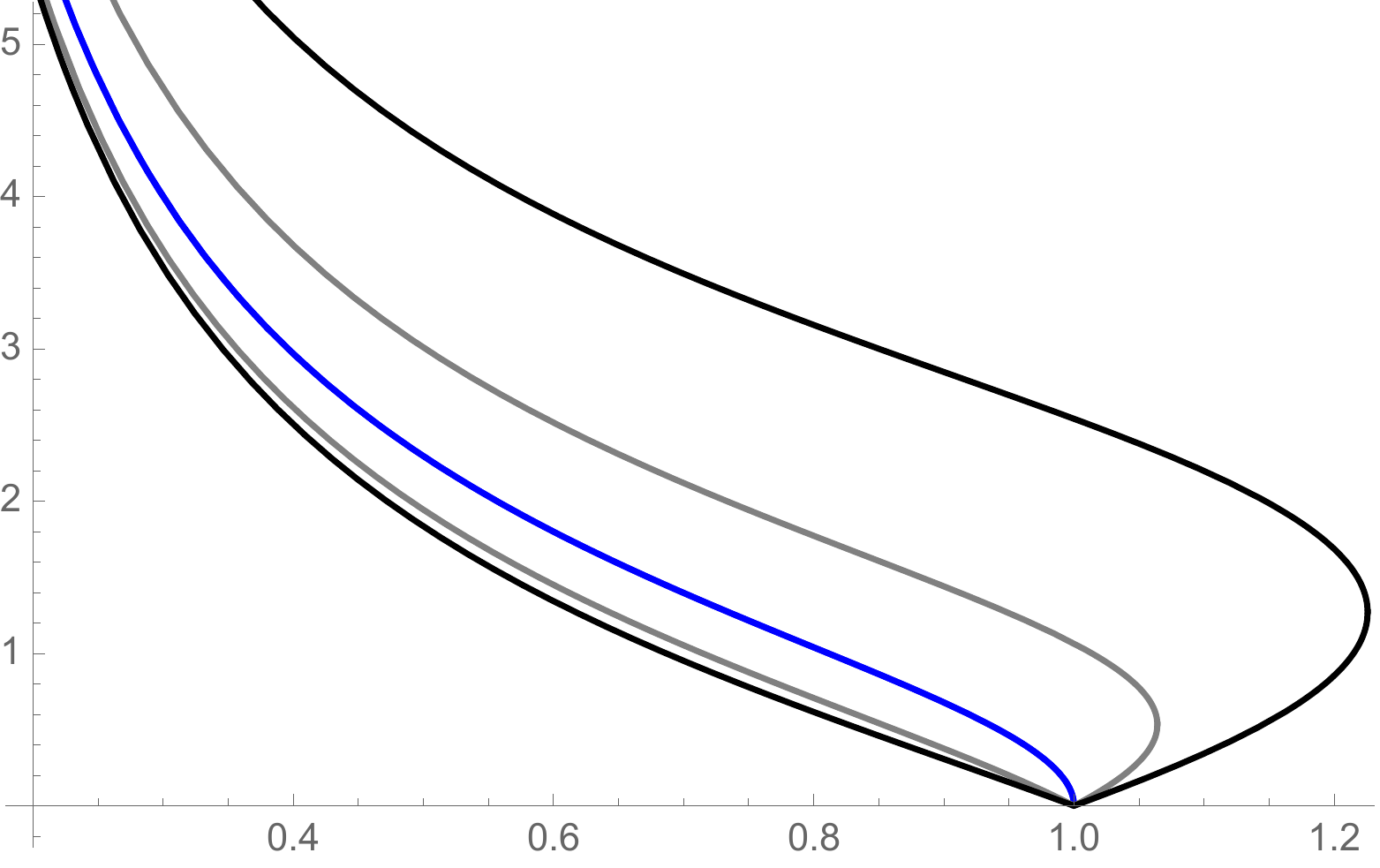}
\end{tabular}
\put( -220,-54) { $x$}  \put(-440,53) { $S^\pm$}
\put( -1,-54) { $x$}  \put(-220,53) { $S^\pm$}
\caption{The entropy inside each mouth $\frac{-v_b }{4\pi b^3\sigma'}S^\pm$. 
Here, $q=2/3$ (L) $q=2$ (R) and $\displaystyle \kappa =0$ (the symmetric solution), $0.2 $ and $0.4$, respectively for blue, gray and black curves.
The wormhole throat is located at $x=1$.
Of the two curves of the same color, $S^\pm$ corresponds to the right/left curve of the blue one, respectively. 
}
\label{fig:tm}
\end{center}
\end{figure} 
The entropy at each side of the handle always increases with $r_\pm$. 
As can be seen from Eq.~\eqref{Kfn1}, $S^-$ diverges as $x \to0$ when $q> 1$.  
On the other hand, it converges to a finite value in the limit when $0<q< 1$. 
Regardless of the value of $q$, the entropy $S^+$ approaches a finite value as $x \to \infty$.

We next consider the case $\kappa \leq 2q^{-1}(1-1/2q)^{2q-1}$.
The symmetric solution $u=x$ belongs to this case and its radius satisfies $y =  x^{-q}$.
Let us consider an asymmetric solution.
The formula for the Right solution is exactly the same as Eqs.~\eqref{r:sol1-1}, \eqref{m:sol1-1}, and \eqref{f:sol2-1} except for the range of $x$ and $u$ are restricted to $0\leq x \leq x_M$ and $u_M\leq u \leq 1$.
Therefore, it is enough to write the Left solution here. 
In Eq.~\eqref{sol2}, the Right solution curve is smoothly connected to the Left curve at $x=x_M$.
The range of $u$ and $x$ for the Right solution curve are $0\leq x\leq x_M$ and $0 \leq u\leq u_M$, respectively. 
The areal radial coordinate becomes
\be{r:sol1}
\frac{r}{b} = y = x^{-q} \times
 \exp \left|2\mathfrak{I}_{q,\kappa}(x_M)-\mathfrak{I}_{q,\kappa} (x) \right| .
\ee
At the present case, both ends of the solution curve $C_\kappa$ are located at $(u,x) =(0,0)$.
As $x\to 0$, the function $\mathfrak{I}_{q,\kappa} (x)$ goes to a constant number. 
Therefore, the radius goes to infinity in the limit forming two asymptotic regions at both ends. 
The mass for the Left solution becomes
\be{m:sol2}
m(x) =  \frac{b}{2}
 \left[x^{1-q}-\frac{\kappa x^q}{2} -\sqrt{\kappa}
 	\sqrt{\frac{\kappa}4 x^{2q}+ 1-x}
	 \right] \exp \left[2\mathfrak{I}_{q,\kappa}(x_M)-\mathfrak{I}_{q,\kappa} (x) \, \right]   .
\ee
The mass function goes to infinity or approaches a finite value when $q> 1$ or $0<q<1$, respectively.
The metric function $f_+(r)$ for the Left solution becomes
\be{f:sol2}
f_+(r) = f_b \times
  \exp \left[-4\mathfrak{I}_{q,\kappa}(x_M)+2\mathfrak{I}_{q,\kappa} (x) \right]  .
\ee

The entropy $S^-$ for the bottom half is given by the same formula as Eq.~\eqref{S:sol2}.
On the other hand, to calculate $S^+$, we need to consider both the Right and the Left solution curves depending on the value of $u$ relative to $u_M$.
We should first integrate $L_+$ along the Right curve from $u=1$ to $u_M$ and then continue to integrate it along the Left curve. 
Therefore, $S^+$ becomes
\be{S+:sol2}
S^+=\int_b^{r_+} L_+ dr = \frac{4\pi b^3 \sigma'}{-v_b} \left[ \theta(u_+-u_M) K_{q,\kappa}(x_+) 
+\theta(u_M-u_+) \Big(2K_{q,\kappa}(x_M)- K_{q,\kappa} (x_+) \Big) \right].
\ee
The entropy $S^+$ increases with $r_+$. 
As can be seen from Eq.~\eqref{Kfn1}, $S^+$ diverges as $x \to0$ when $q> 1$.  
When $0< q< 1$, $S^+$ approaches a constant value in the limit.
This behavior of entropy is consistent with that of mass in Eq.~\eqref{m:sol2}.

\subsection{Thinshell wormhole case}
Consider a thin-shell wormhole in which matter is localized around the throat, i.e., $r_\pm \sim b$.
See Ref.~\cite{Israel:1966rt} for its construction.
A nearby solution around the throat is enough to describe the geometry.
For this purpose, we introduce a new radial coordinate
$$
z = 1-x.
$$

The zeroth order solution is nothing but $\kappa(1-u) = z^2$ in Eq.~\eqref{0th}. 
To calculate the entropy of the wormhole, we need a solution an order higher. 
Considering a power law solution around $z=0$, we find the next order solution of  Eq.~\eqref{TOV},
\be{sol4}
1-u = \kappa^{-1} z^2\big[1 + 2 \left( q + s{\kappa^{-1}} \right) z + \cdots \big].
\ee
Using Eqs.~\eqref{sol4} and~\eqref{xi:r}, the radius and the mass satisfy
\be{y:z}
y = 1+ \frac{q}{\kappa} z^2 +\cdots, 
\qquad 
m(r) = \frac{b y u}{2} \approx \frac{b}{2}\Big(1+\frac{q-1}{\kappa} z^2 +\cdots \Big).
\ee

Using these results, the entropy~\eqref{S WH} in the upper/lower half of the wormhole becomes 
\be{S:ts}
S^\pm
=\frac{4\pi b^3 \sigma' }{- v_b\sqrt{\kappa} }   |z| .
\ee
Therefore,  the total entropy of a wormhole is 
\be{S:thinshell}
S_{\rm WH} = \frac{4\pi b^3 \sigma' }{- v_b\sqrt{\kappa} }( |z_+|+ |z_-|) ;
\qquad z_\pm = \frac{v_\pm}{v_b} -1.
\ee
Note that the proper-radial length at the throat is 
$\Delta  \equiv \delta r/\sqrt{1-u} \approx 2bq/\sqrt{\kappa} \times \delta z$ where $\delta z=|z|$.
Putting this to the entropy formula, we get the entropy to be
$$
S_{\rm WH} \approx \frac{\sigma'}{-2v_b q} \times V; \qquad V = 4\pi b^2 (\Delta_++\Delta_-),
$$
where $V$ is nothing but the volume of the handle when $|z_\pm| \ll 1$.
This result is clearly consistent with the entropy density in Eq.~\eqref{sA}. 

\subsection{Entropy of a regular asymptotic wormhole}

In most cases, the matter in a wormhole is assumed to be restricted to a narrow region around the throat or at least in a handle. 
But even if matter spreads to infinity to form regular asymptotic regions, it is physically sound.
In the case of a wormhole spacetime consisting of a single anisotropic matter, it was shown in Ref.~\cite{Kim:2019ojs} that a regular asymptotic region exists only when $0< \gamma < 1$ where $\gamma \equiv -w_1/2w_2$.  
The asymptotic behavior of the wormhole solutions happens around $(u,x) = (0,0)$ and is given by
\be{u:asym}
 u =  -\frac{2v_b\gamma }{1-\gamma} x + q_\pm x^\gamma ; \qquad x \ll 1,
\ee
where $q_\pm$ is a constant representing asymptotic property of the upper/lower half. 
In general, it must be related with the asymmetry $\kappa$.
Explicitly, for the solution in Sec.~\ref{sec:IVA},  $\gamma=2q$ and $q_-=(1-2q)^{-1}-\kappa$. $q_+$ is not defined because the solution does not have an asymptotically free region.
For the solution in Sec.~\ref{sec:IVB}, $\gamma=q$ and $q_\pm = \pm \sqrt{\kappa}$. 

The radius and the mass function take the form,
\be{rm:asym}
r = b_\pm x^{-\gamma}, \qquad 
m_\pm(r) = \frac{ur}{2} = \frac{b_\pm}{2} \left[q_\pm - \frac{2\gamma v_b}{1-\gamma} \left(\frac{r}{b_\pm}\right)^{(\gamma-1)/\gamma}\right]. 
\ee
where $b_\pm$ is a length scale which must be related to $b$. 
Explicitly, for the exact solution with $s= 0$ in Sec.~\ref{sec:IVA}, $b_- = b/\kappa$.
We cannot identify $b_+$ because there is no asymptotic region.
For the exact solution with $s=-1$ in Sec.~\ref{sec:IVB}, for the case $\kappa< 2q^{-1} (1-1/2q)^{2q-1}$, 
\be{al:sol2}
b_- = b
  \exp \left|\mathfrak{I}_{q,\kappa} (0) \right|  \quad \mbox{the Right}  , \qquad  
b_+ = b\exp \left|2\mathfrak{I}_{q,\kappa}(x_M)-\mathfrak{I}_{q,\kappa} (0) \right|   
   \quad   \mbox{the Left} .
\ee
For the case $\kappa> 2q^{-1} (1-1/2q)^{2q-1}$, $b_-$ is given by the same equation but $b_+$ is ill-defined.  

The ADM mass of the wormhole in the upper/lower half becomes
$$
M_\pm = \lim_{r\to \infty} m_\pm (r) = \frac{b_\pm q_\pm}{2} .
$$
The metric function~\eqref{f pm} takes the form 
\be{fpm}
f_\pm(r)= \alpha_\pm^2 \Big(1- \frac{2M_\pm}{r} +\cdots\Big); \qquad 
\alpha_\pm = \sqrt{f_b} \left(\frac{b_\pm}{b}\right)^{-\frac{q}{\gamma}}.
\ee

The entropy inside $r$ of the upper/lower half of the wormhole is
\be{S:asym}
S^\pm = \frac{4\pi b^3 \sigma'\sqrt{\kappa} }{ v_b (s+1)}
	\left[  \frac{1}{ \sqrt{\kappa}} \left(\frac{b_\pm}{b}\right)^{1-\frac{q}{\gamma}} 
	\frac{q_\pm-\frac{s+1}{2q+1-s} x^{1-\gamma} }{\sqrt{1-q_\pm x^\gamma -\frac{2\gamma v_b}{\gamma-1} x }} \pm 1  \right].
\ee
Therefore, the total entropy of the thin-shell wormhole becomes
\be{S:asym}
S_{\rm WH} = \frac{8\pi b^2 \sigma' }{ v_b (s+1)}
	\left[M_+\left(\frac{b_+}{b}\right)^{-\frac{q}{\gamma}} 
	\frac{1-\frac{s+1}{2q+1-s}q_+^{-1} x_+^{1-\gamma} }
		{\sqrt{1-q_+ x_+^\gamma -\frac{2\gamma v_b}{\gamma-1} x_+ }} +
	M_- \left(\frac{b_-}{b}\right)^{-\frac{q}{\gamma}} 
	\frac{1-\frac{s+1}{2q+1-s} q_-^{-1} x_-^{1-\gamma} }{\sqrt{1-q_- x_-^\gamma -\frac{2\gamma v_b}{\gamma-1} x_- }}  \right].
\ee
In the $r\to \infty $ limit, the entropy approaches a finite value 
\bea
\lim_{r\to \infty} S_{\rm WH} &=& 
		\frac{8\pi b^2 \sigma' }{ v_b (s+1)}
	\left[M_+\left(\frac{b_+}{b}\right)^{-\frac{q}{\gamma}}
	+M_- \left(\frac{b_-}{b}\right)^{-\frac{q}{\gamma}} \right] \nnl
&=&  \frac{4\pi b^2 \sigma'}{ v_b \sqrt{f_b} (s+1) }
	\big(\alpha_+ M_+
	+\alpha_-M_- \big) . \label{S:infty}
\eea
Note that $\sigma'/(-v_b)$ determines the signature of the entropy density. 
Therefore, to have a positive definite entropy, the signature of $s+1=2(1+w_1+2w_2)/(1+w_1)$ should be different from that of 
\be{M:gen}
\mathcal{M} \equiv \alpha_+ M_++\alpha_-M_- = \frac{b\sqrt{f_b}}{2}
	\left[\Big(\frac{b_+}{b}\Big)^{1-q/\gamma} q_+ +
	\Big(\frac{b_-}{b}\Big)^{1-q/\gamma} q_- \right] .
\ee
This presents a global constraint on the ADM masses of a wormhole, given by what constitutes the wormhole. 
For the solution with $s=-1$ in Sec.~\ref{sec:IVB},  this constraint determines $\mathcal{M} = 0$ for every solutions having two asymptotic regions in which matter spreads over.
  
\section{Summary and discussions} \label{sec:V}

We have obtained the entropy of a static spherically symmetric wormhole based on the assumption that the entropy comes from the matter entropy constituting the wormhole.
The entropy density of anisotropic matter having linear equation of state, $p_i = w_i \rho$, was given as a function of density and radius.   
We integrated the entropy density over spatial volume of the wormhole to get the total entropy.
Wormhole solutions were expressed as integral curves on two dimensional $(u,x)$ plane where $u=2m(r)/r$ and $x= -8\pi w_1 r^2 \rho$, respectively.
A solution curve smoothly connects the boundary data $(u_+=2M_+/r_+, x_+= -8\pi w_1 r_+^2\rho_+)$ at $r _ +$ in the upper half of the wormhole with those at $ r _- $ in the lower half through the throat at $(u,x)=(1,1)$. 
We have found that the entropy of a wormhole is given by, for $1+w_1+2w_2\neq 0$, 
$$
S_{\rm WH} = S^++ S^-; \qquad 
S^\pm  = \frac{-8\pi b^3 \sigma' w_1}{s+1}  \left[ 
	\frac{ u_\pm -x_\pm  }{\sqrt{1-u_\pm }} x_\pm^{-q } 
		y_\pm^{\frac{s+1}{2} }  \mp \sqrt{\kappa} \right] , \quad y_\pm \equiv \frac{r_\pm}{b}, 
$$
where $\kappa$ represents the asymmetry between the upper and the lower halves and $q, s$ are constants determined from the equation of state given in Eq.~\eqref{vb}. 
Even though this is a general formula, it contains many dummy data come from the mouths at $r=r_\pm$ and from the throat at $r=b$.
To remove the dummies, one should put the exact solution curve and the radial dependence. 

Explicitly, we have shown three exact entropy forms for $1+w_1+2w_2=0$, $1+w_1+4w_2=0$, and the near-throat solutions. 
Based on the exact solutions, simplified forms of the wormhole entropies were presented.
We also displayed the entropy for the asymptotic solutions. 
The entropy of the asymptotic limit is proportional to $\mathcal{M}\equiv \alpha_+ M_+ + \alpha_- M_-$ where $\displaystyle \alpha_\pm \equiv \lim_{r\to \infty} \sqrt{-g_{tt}}$ in the upper/lower half, respectively.
An interesting conclusion is that the signature of $\mathcal{M}$ is fully determined by the signature of $(1+w_1+2w_2)/(1+w_1)$.
Explicitly, $\mathcal{M}=0$ for every regular wormhole solutions consisting of matter satisfying $1+w_1+2w_2=0$, where the matter is allowed to spread over the whole space.
This strongly suggests that $\mathcal{M}$ could be an important physical parameter for the wormhole system representing a total energy. 

In Ref.~\cite{Kim:2019elt}, the thermal equilibrium was defined unambiguously based on the generalized temperature in Eq.~\eqref{Tgen1}.
Considering a volume preserving (here  we simply assume $r_\pm$ does not change) process of heat flow through the mouths of a wormhole, 
the first law of thermodynamics can be written as
$$
\delta S_{\rm WH} = T_+^{-1} \delta M_+ + T_- ^{-1}\delta M_-= \mathfrak{T}^{-1} \delta\mathcal{M},
$$
where $\mathfrak{T}$ is the generalized temperature and $T_\pm$ is the asymptotic temperatures of the wormhole at the upper/lower half, respectively.
This equation also present the total energy to be the same combination of ADM masses.

So far we have assumed that the wormhole is an inter-universe one.
Even if one considers an intra-universe one, entropy calculations may not be different if the matter is confined to the handle. 
But if the matter spreads throughout the entire space, forming a regular asymptotic wormhole, some discussion are required.  
The spacetime is locally static but not globally static in that the metric component $g_{tt}(\infty)$ may not be uniquely defined. 
Then, the Tolman temperature must be ambiguous. 
Even with all of these troubles, thermal equilibrium can be defined unambiguously based on the generalized temperature.
Thus, we expect that the same form of thermodynamic law is applicable. 
Still, we need to clarify the wormhole temperature based on the heat flow which consider the geometrical back-reaction of a wormhole.
This is the future direction of our research.

\section*{Acknowledgment}
This work was supported by the National Research Foundation of Korea grants funded by the Korea government NRF-2017R1A2B4008513.

\end{document}